\documentclass{article}

\usepackage[english]{babel}

\usepackage[letterpaper,top=2cm,bottom=2cm,left=3cm,right=3cm,marginparwidth=1.75cm]{geometry}

\usepackage{amsmath}
\usepackage{graphicx}
\usepackage[colorlinks=true, allcolors=blue]{hyperref}

\title{Evolution of virial clouds-I: from surface of last scattering up to the formation of population-III stars}

\author{Noraiz Tahir$^{1,2,*}$, Asghar Qadir$^{3,**}$, Muhammad Sakhi$^{4,\dagger}$, and Francesco De Paolis $^{1,2,\ddagger}$ \\~\\

$^1$ Department of Mathematics and Physics ``E. De Giorgi'', University of Salento, \\
Via per Arnesano, I-73100  Lecce, Italy.\\
$^2$ INFN, Sezione di Lecce, Via per Arnesano, I-73100 Lecce, Italy.\\
$^3$ Abdus Salam School of Mathematical Sciences, Government College University.\\
$^4$ Department of Physics, Quaid.e.Azam University, Islamabad, Pakistan.\\~\\

$*$ noraiz.tahir@le.infn.it\\
$**$ asgharqadir46@gmail.com\\
$\dagger$ sakhi.cosmos@gmail.com\\
$\ddagger$ francesco.depaolis@le.infn.it}

\begin{document}
\maketitle

\begin{abstract}
The analysis of WMAP and  {\it Planck} CMB  data has shown the presence of temperature asymmetries towards the halos of several galaxies, which is probably due to the rotation of clouds present in these halos about the rotational axis of the galaxies. It had been proposed that these are hydrogen clouds that {\it should} be in equilibrium with the CMB. However, standard theory did not allow equilibrium of such clouds at the very low CMB temperature, but it was recently shown that the equilibrium {\it could} be stable. This still does not prove that the cloud concentration and that the observed temperature asymmetry {\it is} due to clouds in equilibrium with the CMB. To investigate the matter further, it would be necessary to trace the evolution of such clouds, which we call ``virial clouds'', from their formation epoch to the present, so as to confront the model with the observational data. The task is to be done in two steps: (1) from the cloud formation before the formation of first generation of stars; (2) from that time to the present. In this paper we deal with the first step leaving the second {\bf one} to a subsequent analysis.
\end{abstract}


\section{Introduction}
It had been proposed that a fraction $f$ in the form of molecular hydrogen clouds at exactly the CMB temperature in galactic halos may be a significant component of their dark
matter \cite{1}. To look for these \textit{chameleon} clouds one suggestion was to look for an asymmetric Doppler shift in the CMB, due to the rotation of these clouds in the M31 halo \cite{2}. The shift was observed first by analyzing WMAP \cite{3} and then, more precisely, the {\it Planck} data \cite{4}. This effect opened up a window to study galactic halos in more detail, and was thought to be a significant method to constrain the missing baryons \cite{5}. In a series of papers the CMB data was used to trace the rotation of the halos of some nearby galaxies \cite{6,7,8,9,10}. The data showed a temperature antisymmetry not only of the galactic disks but also of the galactic halos up to distances of hundreds of kpc from the galactic center, and the asymmetry was also used to study the rotational dynamics of the halos of some nearby spiral galaxies \cite{11,12}. The fact that the temperature asymmetry was almost frequency independent strongly supports the Doppler shift explanation to observe the chameleon clouds in the halos. However, this did not prove that the clouds are made by pure molecular hydrogen, as originally proposed, since helium, dust and other heavier molecules may also be present, and they might contaminate these clouds with a significant fraction. These clouds were to survive on account of the virial temperature being in equilibrium with the CMB, we called them ``virial clouds'' \cite{13}. It is suggested  that no such equilibrium could exist at the current, very low, CMB temperatures, as there would be no mode that could be excited by the photons and the available clouds might collapse to form stars or other planetary objects \cite{13a}. Recently it was demonstrated that this equilibrium {\it does} arise on account of the
translational mode, despite the extremely small probability, because of the
size of the virial clouds and the time scales available for thermal equilibrium to be reached, so that the time required for thermalization is much less than that required for collapse \cite{14}.

Proving that virial clouds {\it can} exist does not prove that they {\it do} 
exist and play a significant role in the observed temperature asymmetry in 
the CMB towards some galaxies. To take the matter further we need to trace 
the evolution of these virial clouds from the time of their formation to the present, allowing for all the major qualitative changes that occurred during that time. 

Virial clouds would have been formed at the time of the {\it surface} of last scattering (SLS) at $z=1100$ when the CMB temperature was about 3000 K. At this stage recombination occurred and the CMB filled the Universe with a red, uniformly bright glow of blackbody radiation. With the passage of time the temperature decreased and the CMB shifted to the infrared and finally to the microwave region of the electromagnetic spectrum. To human eyes, the Universe would then have appeared as a completely dark place. A long period of time had to pass until the first objects collapsed, forming the population-III that shone in the universe with the first light ever emitted that was not part of the CMB \cite{barkana2001}. In order to maintain the stability virial clouds {\it must} be in quasi-static equilibrium with the CMB since SLS. Obviously, when the clouds formed their chemical composition was the primordial one with about 75\% atomic hydrogen and the rest helium. Afterwards, we expect a fast change of their composition as a consequence of the formation of population III stars. The question that arises is ``When did the population-III form?'' Because the primordial density fluctuations in the Universe are random, the question of when the very first star formed does not have a simple answer. The time when the first halo collapsed depends on how rare a fluctuation we are willing to consider. For example a $5\sigma$ fluctuation in the density field can lead to the collapse of the halo and the formation of population-III at $z=30$ \cite{darkage}. Using this large volume the population-III in the sky should be one formed from $8\sigma$ fluctuations at $z= 48$ when the temperature of the Universe was about 132 K \cite{Yoshida_2004}. Because of the qualitative changes that took place during and after the formation of population-III, we intend to model the evolution of these virial clouds in two steps: (1) from SLS at $z=1100$ when the temperature of the Universe was 3000 K, up to the formation of population-III at $z=50$ when the temperature of the Universe was $137$ K; (2) from $z=50$ when the population-III were forming, and consequently the composition of the interstellar medium (ISM) changed \cite{pop3} (and hence the clouds), to the present. It must be admitted that there would be significant changes during the second step, so it will be studied more clearly and in more detail later on.

In this paper we study the first step of the evolution of virial clouds, which begin at SLS about 380,000 years after the Big Bang at $z=1100$ when the Universe was very hot $\sim 3000$ K, up to the formation of population-III at $z=50$ when the Universe was $137$ K. Since at the SLS the Universe consists of primordial abundance of atomic hydrogen and helium i.e. $\sim 75\%$ H, and $\sim25\%$ He by mass, this ratio should have remained almost the same till the formation of population III stars \cite{barkana2001}. There were other molecules and atoms that would have
been present at that time, which include deuterium, helium-3, lithium and
molecular hydrogen \cite{peebles, wagoner}, which could have contributed to
the virial clouds. Molecular hydrogen played a role as a coolant in the
clouds \cite{peeblesa,saw}, and it would have formed in traces inside the
clouds in the early stages. During the time span under study the ratio
of molecular hydrogen to atomic hydrogen $n_{H_2}/n_H$ is $\approx 10^{-6}$
\cite{Friedman}. Moreover, the ratio of other elements like deuterium,
helium-3, and lithium to atomic hydrogen is negligible \cite{lepp}. The ratio
of atomic hydrogen and helium would remain the same and these would be the
main component to form the virial clouds during the period.

The plan of the paper is as follows: In Section \ref{distribution} we will
use the canonical distribution for obtaining the changing density, size and
mass of a single fluid virial cloud from z=1100 to z=50 (see Section \ref{singlefluid}). In the next section we extend our analysis to the two fluids model with He {\it and} H in the primordial mix (Section \ref{two}). Finally, in Section \ref{results} we conclude with a discussion of the results.
\section{\label{distribution} Canonical Distribution Function}
Since the clouds must be considered to be in thermal equilibrium because they
are embedded in the heat bath of the CMB \cite{18}, we need to use the
canonical distribution function for a fixed temperature and use the cooling
of the heat bath to provide a quasi-equilibrium. This is
justified because the changes are very slow compared to the scale of
thermalization. One may wonder if quantum effects may not be significant and could affect the physical parameters of virial clouds. Since the present paper deals with temperatures above 137 K, quantum effects are negligible, accounting for the cloud density. We will start by considering single fluid clouds and then go on to the  two fluids case. Since the gas to be considered is either H or He, it will be monatomic. For the  monatomic gas the Hamiltonian incorporates only the translational mode. The clouds should start to form in the potential well of cold dark matter (CDM). As the clouds are thermalized the potential well will not cause them to collapse to form population-III stars \cite{barkana2001}, but will modify the physical parameters i.e. the mass, radius and central density.
\subsection{\label{singlefluid} Single Fluid}
For the sake of simplicity, we start our analysis by considering pure atomic
hydrogen clouds. Since atomic hydrogen is monatomic, we can easily get the
results by using the density distribution and Lane-Emden equation (obtained
in \cite{14}) with the boundary condition that the density is exactly zero at
the edge of the cloud. The equations are 
\begin{equation}
	\rho(r)=8m^{5/2}\left(\frac{G\rho_{c}}{3k_BT}\right)^{3/2}
	exp\left(-\frac{GM(r)m}{rk_B T}\right),
	\label{e1}
\end{equation}
and
\begin{equation}
	r\frac{d\rho(r)}{dr}-r^{2}\left(\frac{4\pi Gm}{k_B T}\right)\rho^{2}(r)-\rho(r)\ln\left(\frac{\rho(r)}{B}\right)=0,
	\label{e2}
\end{equation}
where, $B=(8m^{5/2}/3^{3/2})(G\rho_c/k_B~T)^{3/2}$, $m$ is the mass of hydrogen atom, $1.67 x 10^{=27} kg$, $G$ is Newton's gravitational constant, $\rho_c$ is the central density of the cloud, and $M(r)=\int_{0}^{r}4\pi\rho(q)q^{2}~dq$ is the mass interior to $r$.

\begin{figure}
\centering
\includegraphics[width=0.6\textwidth]{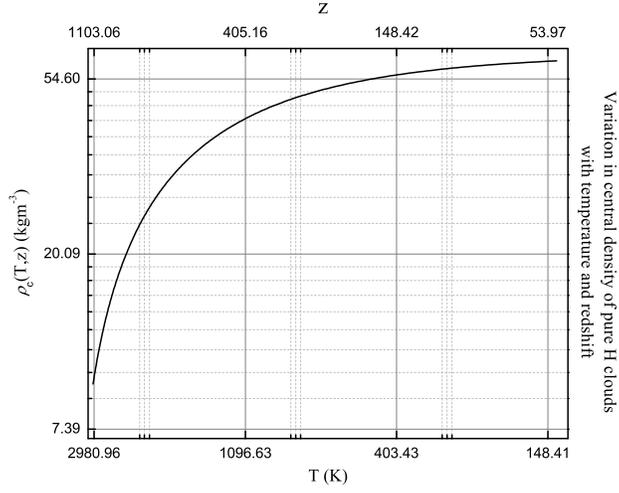}
\caption{\label{fig1}Variation in the  central density $\rho_{c}$ of the virial clouds with respect to both the red shift (indicated by the upper horizontal axis) and  the CMB temperature (bottom axis) for pure H clouds.}
\end{figure}

The above expressions gives us the central density of the cloud. Now we need
the Jeans mass and radius. From the virial theorem, $2K+\Phi=0$, where, $K$
is the kinetic energy of the molecules and $\Phi$ is the gravitational
potential, the Jeans mass squared is \cite{chandra}
\begin{equation}
	M_{J}^2 \simeq \left(\frac{81}{32\pi\rho_{c}}\right)\left(\frac{k_BT}{Gm}\right)^3,
	\label{1}
\end{equation}
and the corresponding Jeans length (radius) squared is
\begin{equation}
	R_{J}^{2}=\frac{27k_BT}{20\pi\rho_{c}Gm}.
	\label{2}
\end{equation}
We use eq.(\ref{e2}) to estimate the central density of the
clouds. We solve the equation numerically with a guess value of $\rho_{c}$,
at a fixed temperature, and see where the density becomes exactly zero at the
boundary. We then compare the Jeans radius with that central density with the
value available to us. We adjust the central density so that the density
becomes zero exactly at the Jeans radius. In this way we get a
self-consistent solution of the differential equation subject to the given
boundary conditions. Next, we decrease the temperature and repeat the
process. The evolution of  the central density of these clouds as a function of
both the time and  the CMB temperature is shown in Fig.\ref{fig1} the density of these clouds tend to increase with the decrease in temperature.
We also give the evolution of the Jeans mass (see Fig.\ref{fig2}) and radius (see Fig.\ref{fig3}) of pure H virial clouds.
\begin{figure}
\centering
\includegraphics[width=0.6\textwidth]{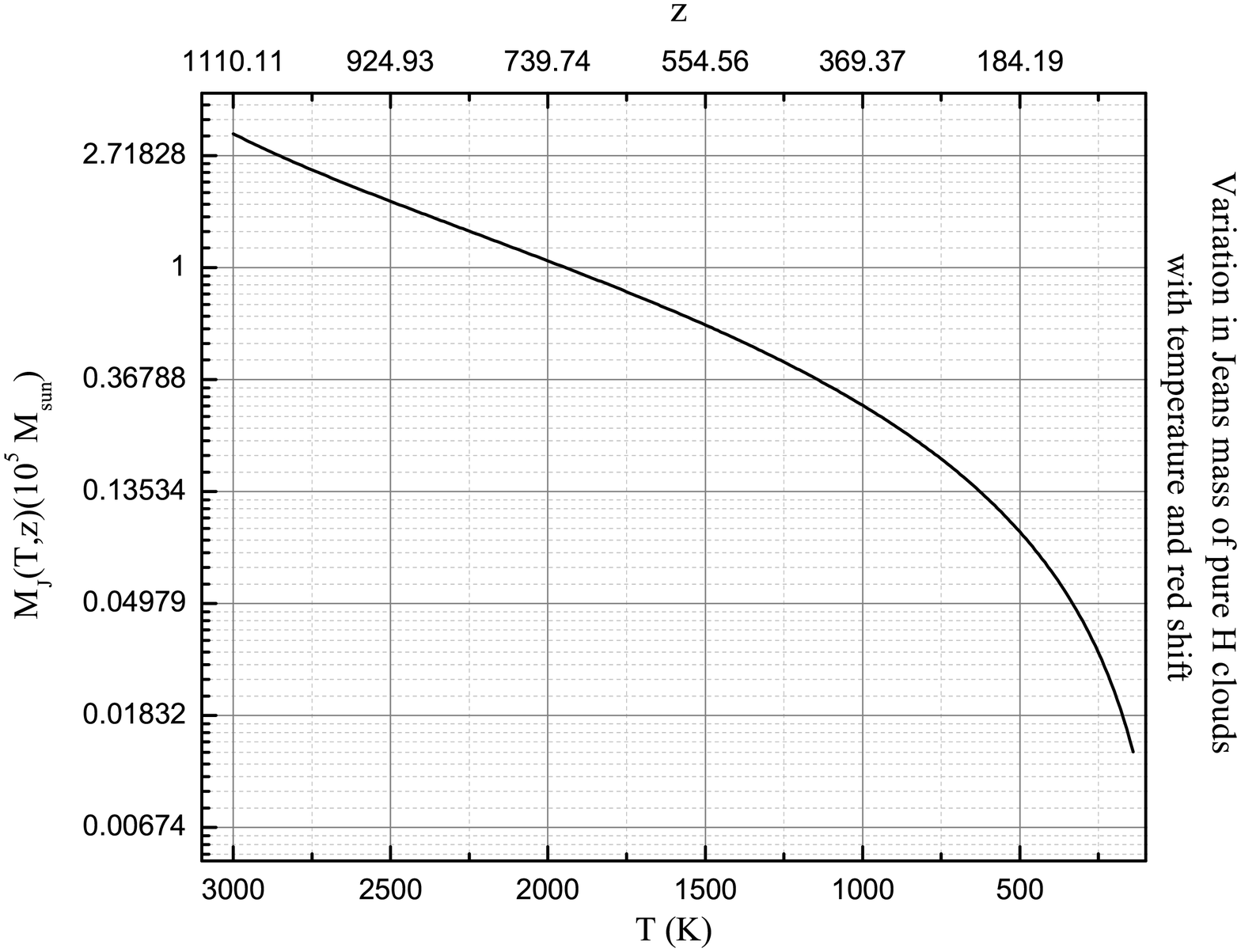}
\caption{\label{fig2}Variation in the Jeans mass of  pure H virial clouds.}
\end{figure}

\begin{figure}
\centering
\includegraphics[width=0.6\textwidth]{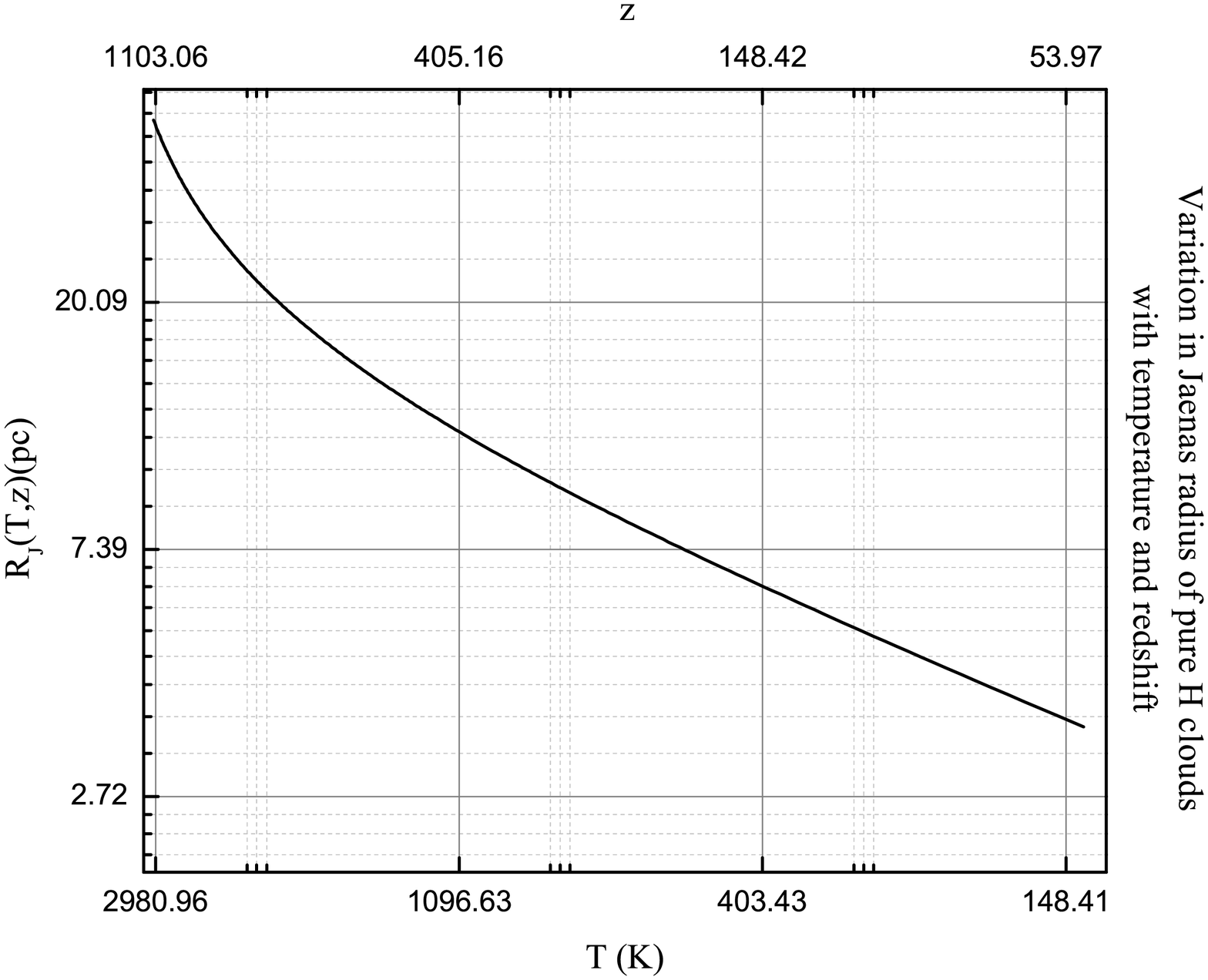}
\caption{\label{fig2}Variation in the Jeans radius of  pure H virial clouds.}
\end{figure}

\subsection{\label{two} Two-Fluid Virial Cloud}

For generality we consider a virial cloud composed of an arbitrary mixture of H and He, with mass fractions $\alpha$ and $\beta$. Then, we use the primordial cosmological  fractions of H and He for the final computation. The total mass of the cloud is, obviously,  $M_{cl}(r)=\alpha M_{H}(r)+\beta M_{He}(r)$,  with the condition $\alpha+\beta=1$. The density distribution for two fluids is given by \cite{14}
\begin{align}
	\rho_{cl}(r)=\sqrt{\frac{64}{27}}\frac{(G\rho_{c_H}\rho_{c_{He}})^{3/2}}{(kT_{CMB})^{9/2}}(m_Hm_{He})^{5/2} exp\left[-\frac{1}{2}\left(\frac{\alpha G M_H(r)m_H}{rk_BT}+\frac{\beta G M_{He}(r)m_{He}}{rk_BT}\right)\right].
\end{align}
and the corresponding Lane-Emden equation is
\begin{align}
	r\frac{d\rho_{cl}(r)}{dr}-r^2\left(\frac{2\pi G}{k_BT}\right)[\rho_{cl}(r)(\alpha \rho_{c_H}m_H +\beta \rho_{c_{He}}m_{He})]-\rho_{cl}(r)\ln\left(\frac{\rho_{cl}(r)}{\tau}\right)=0,
\end{align}
where,
$\tau=(8/3\sqrt{3})[(G\rho_{c_H}\rho_{c_{He}})^{3/2}/(k_BT)^{9/2}][m_Hm_{He}]^{5/2}$,
$\rho_{c_{H}}$ is the central density of H cloud, $\rho_{c_{He}}$ is the
central density of He cloud, $m_H$, the mass of single atom of hydrogen, and
$m_{He}$, the mass of single atom of helium.

Using the same approach we used to solve the Lane-Emden equation for a single
fluid, we estimated the central density, Jeans mass and radius of the
two-fluid virial clouds with primordial fraction of H, and He, i.e. $\alpha=
0.75$, and $\beta=0.25$.

\begin{figure}
\centering
\includegraphics[width=0.6\textwidth]{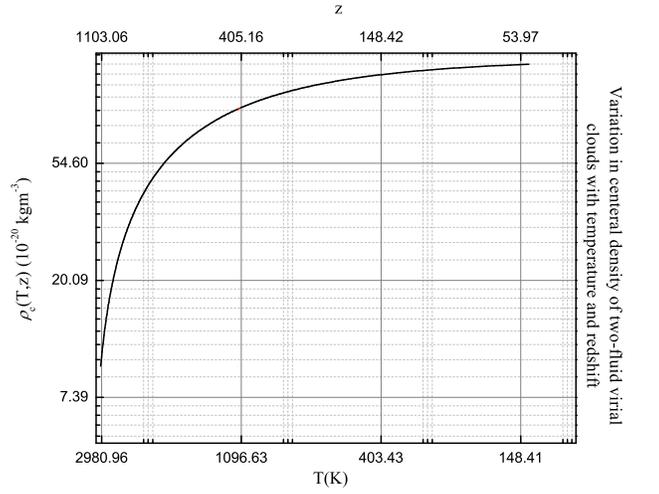}
\caption{\label{fig7}Variation in the central density of the virial clouds  made up by a mixture of  H and He with cosmological primordial relative abundances.}
\end{figure}

\begin{figure}
\centering
\includegraphics[width=0.6\textwidth]{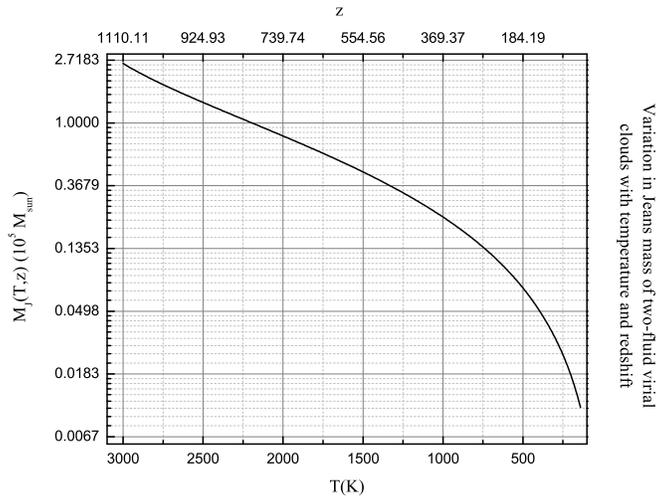}
\caption{\label{fig8}Variation in the  Jeans mass of the virial clouds constituted by cosmological primordial fraction of H and He.}
\end{figure}

\begin{figure}
\centering
\includegraphics[width=0.6\textwidth]{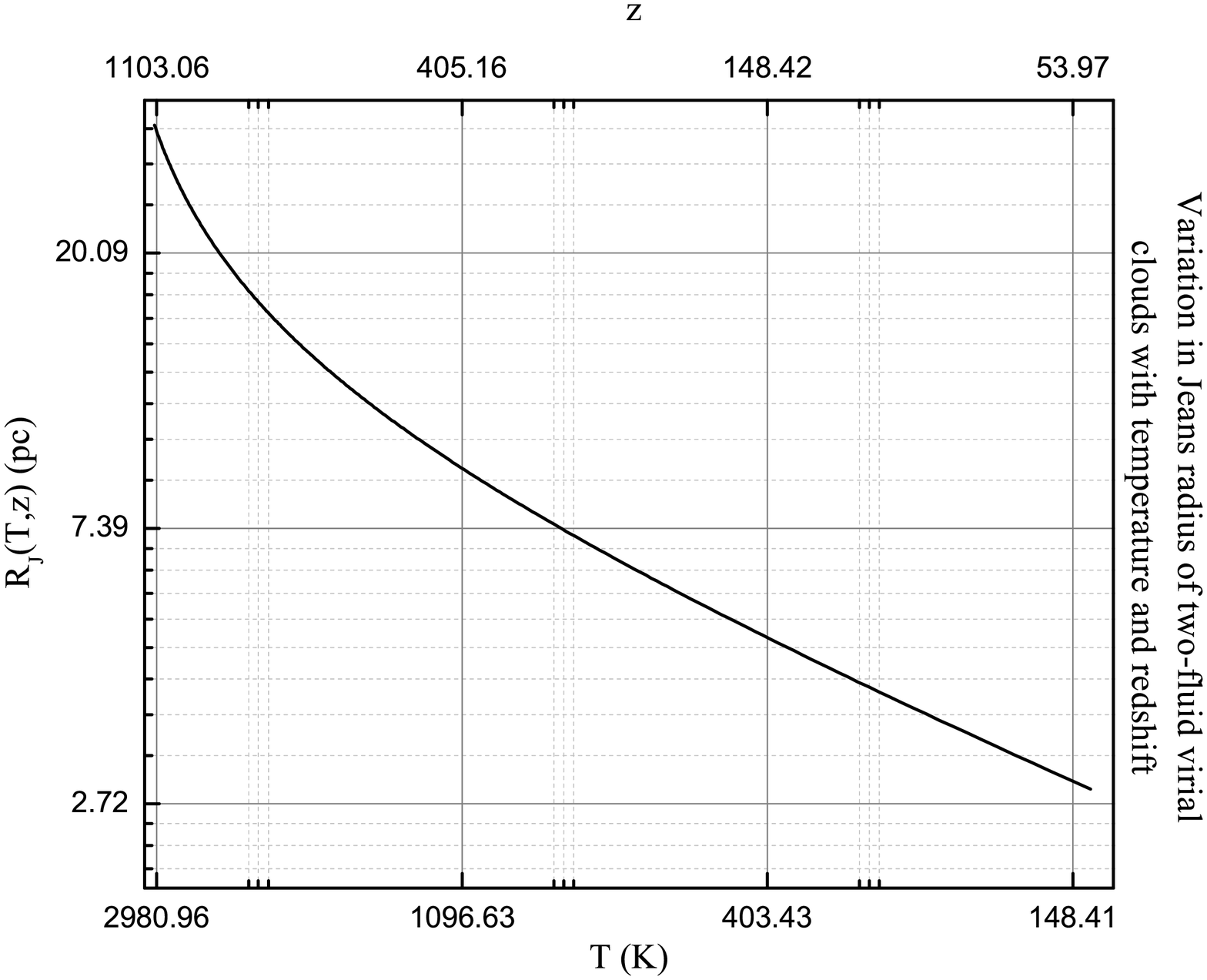}
\caption{\label{fig9}Variation in the Jeans radius of the virial clouds having primordial fraction of H and He.}
\end{figure}

The evolution of the  central density of the two-fluid H-He virial cloud, both with respect to the time and the CMB temperature, is shown in Fig.\ref{fig7}.
Similarly, the evolution of the  Jeans mass is shown in Fig.\ref{fig8} while the
evolution of  Jeans radius is shown in Fig.\ref{fig9}. As one can see, the physical parameters of the two-fluid clouds do not change substantially with respect to the  pure H clouds.
\section{\label{results} Results and Discussion}

In Section II it has been shown that the  evolution of the virial clouds is {\it precisely} determinable from the SLS up to the formation of population-III stars. In Figs.1-3 we saw that pure H clouds would have been more massive, larger and less dense in the beginning, and would have lost their mass and size as the temperature
decreased, in such a way that their density increased. In the case of clouds constituted by  two fluids with a cosmological primordial mixture of H and He,  the evolution of the cloud  physical parameters does not change substantially with respect to the single fluid model: the cloud central density tends to increase while the Jeans mass and the Jeans radius decrease with the CMB temperature (see Figs. 4-6).   However, with a more detailed analysis of Figs.  4-6 with respect to Figs. 1-3, one can note that  the addition of  He in the clouds makes them slightly denser,  smaller and less massive with respect to the pure H cloud. This is because of the stronger gravitational force for the heavier molecules, but the average effect in the parameters is nearly the same as the single fluid H virial clouds.

Since we are dealing with an era from z=1100 to z=50, one might expect that there would be gravitational clumping in the dark matter gravitational potential well, where there will be gravitational binding energy and hence cooling, so that molecular hydrogen would be more likely to form. The $H_2$ component would be negligible in any case, but more so as there would be no nucleating particles for it in the absence of dust grains to speed up the formation of the molecules.  If any $H_2$ molecule formed during this era it was dissociated by the radiation until the density low enough for them to be stable \cite{saslaw1967}. Hence we do not need to consider the effect of cooling from molecular hydrogen till population-III stars exploded. 

It might be thought that quantum calculations for the scattering cross-section of photon molecule interaction need to be incorporated. We checked the electronic transitions of hydrogen atoms due to CMB photons at $3000$ K. Most of the wavelengths in the {\it Planck} spectrum at $3000$ K do not have enough energy to excite the electron of an H atom from its ground state, but the higher energy tail does have sufficient energy to do so. The percentage of such photons is low but it is not {\it zero}. The energy of the second orbit of hydrogen is $3.41$ eV, so the CMB photons with the same energy or higher can take part in the transition. The total number density of photons at $3000$ K is $5.47\times 10^{17} m^{-3}$ and the number density of photons between 0 to 365 nm at the same temperature is $8.96 \times 10^{13} m^{-3}$, hence the percentage of high energy photon is $0.016$, which is quite low. Hence, there will be a negligible effect of high energy photon which will cause atomic transitions in hydrogen atoms, so we don't need to consider the quantum effects at this stage, but these effects might have a sufficient effect when heavier atoms or molecules contaminate the clouds. These things will be discussed in the second step.

One may wonder if the model presented is not over-simplistic as it ignores any possible rotation of the clouds and any turbulence that may arise or be present in the medium during the formation stage of the virial clouds \cite{Dekel,Rosolowsky}. Our whole purpose in limiting the discussion in this first paper to the time {\it before} population III stars form was to keep the Physics simple without having to neglect some other effects. Unless there is some primordial angular momentum present at the surface of last scattering, there would have to be a mechanism for angular momentum to be generated. The natural mechanism for it to arise in the early stages of galaxy and star formation would be that significant amounts of ejecta from stars, collide with a non-zero impact parameter (as it is most natural to happen), thereby producing local angular momentum. However, that could only occur after the very first stars, i.e. population III stars, explode. This would not only produce the angular momentum that was to be generated locally, it would also produce turbulence and lead to contamination of the clouds by matter that is not primordial. This will have to be investigated in the next paper that will trace the evolution from the formation of population III stars on. For this purpose one would need to estimate how much angular momentum would get generated and whether, and to what extent, it would lead to a rotation of the clouds. Again, one would need to investigate how much turbulence would be generated in the earlier stages of the formation of the clouds, and whether, and to what extent, it would enter into the virial clouds At present it is anticipated that both effects will be minor, and will be amenable to inclusion as small corrections on the basic model. Some effects, such as the inclusion of higher modes in the virial clouds, which would need to use quantum calculations, will occur. Similarly, contamination of the clouds by other matter will occur. Since the mechanism for the new effects is expected to be driven by population III stars, it might be possible to test the model by looking for a correlation between the small-scale anisotropies in the CMB and the distribution of population III stars.

Till the explosion of Population III stars no other fluids or substantial contaminants should be present in the  virial clouds. However, for the subsequent steps of the cloud  evolution we will need to include more fluids (heavier atoms) and dust grains produced by population III stars and ejected into the ISM through their explosions. These dust grains should have an important role in the cloud evolution since they would act as catalysts for the formation of molecular hydrogen \cite{Tinsley}, possibly leading to rapid changes in the cloud  physical parameters, as higher modes would be  excited. In any case, the procedure we have set up in the previous Section  is not limited to the two fluids case and would  be able to accommodate more fluids when the need arises. Of course, this first step  was the simplest one since the composition of the Universe from the SLS to the explosion of Population III stars is precisely known. The Physics involved is also simple as the temperature is comparatively high and quantum effects can be neglected. In the next stage ambiguity arises as the cloud chemical composition becomes more complicated and less well known. However, the Physics is relatively simple. At this stage we {\it will} have to incorporate the quantum calculations that were found to be negligible in the present paper.

\section*{Acknowledgements}
FDP and NT acknowledge the TAsP and Euclid INFN projects. We are most grateful to the anonymous referee for the constructive and useful suggestions, which are particularly relevant for the analysis of the following stages of the virial clouds evolution.

\end{document}